\documentstyle[preprint,aps]{revtex}
\tightenlines
\begin{document}
\draft
\title{Quantum stochastic processes in mesoscopic conductors}
\author{G.J. Milburn}
\address{Cavendish Laboratory, University of Cambridge,\\
Semiconductor Physics Group, Madingley Road, Cambridge, CB3 0HE, UK.\\
 and Department of Physics, The University of Queensland,QLD 4072 Australia.}
\date{\today}
\maketitle

\begin{abstract}
We show an equivalence between the approach of Buttiker and the Fermi quantum stochastic calculus for
mesoscopic systems. To illustrate the method we  derive the current fluctuations in a two
terminal mesoscopic circuit with two tunnel barriers containing a single quasi bound state on the well. The method enables us
to focus on either the incoming/outgoing Fermi fields in the leads, or on the irreversible dynamics of the well state
itself. The quantum stochastic calculus we use is the Fermi analogue of the input/output methods of quantum  optics.
\end{abstract}
\newpage

\section{Introduction}
The theory of conductance in mesoscopic electronics was developed some
years ago by Buttiker, following upon earlier ideas of Landauer\cite{Buttiker}. The conductance of a mesoscopic system is
given in terms of the scattering within and between quantum channels  and involves the transmission and reflection
coefficients as well as the thermal occupation of reservoirs feeding or draining those channels. The theory makes direct
contact with measured currents through averages of quadratic functions of Fermi field operators in the channels. The
computation of the scattering matrices depends on the nature of the systems connected to the reservoirs, which could be a
simple tunnel barrier or an array of coherently coupled quantum dots. In the Buttiker approach once the scattering matrices
are calculated we do not need to refer to the dynamics of any local systems to which the input and output channels are
coupled.  In many ways this theory is a Fermion analogue of the quantum description of optical fields interacting with an
optical cavity under the Markov approximation. Such systems are described by the input/output theory of Collett and
Gardiner\cite{ColGar}. The properties of the fields outside the cavity are determined by a scattering matrix connecting the
input and output fields to the cavity and the dynamics inside the cavity. The input/output theory for optical fields has been
shown to be an example of the quantum stochastic calculus for Boson fields\cite{Barchielli86}. 

The Buttiker approach is particularly useful in determining measured properties of the mesoscopic system, such as conductance.
However recent interest in coherently coupled quantum dots for quantum computation\cite{Kane98,Loss98} has focussed attention
on the dynamics of localised systems, such as quasibound states on quantum dots,  rather than the transport through input and
output channels. In the input/output theory of quantum optics, the dynamics of the local system is described through a Markov
master equation, and there is a consistency between a description entirely in terms of the input and output modes and the
irreversible dynamics of the local system to which they couple. In this paper we establish the connection between a
description in terms of input and output channels and the irreversible dynamics of a localised quasibound state on a single
quantum dot. The analysis is easily extended to more complex local systems. This enables us to make a connection between two
very successful theories, quantum optics and quantum mesoscopics, for the treatment of quantum stochastic processes.  We expect
that analogies between quantum optics and mesoscopic electronics will prove useful as the latter explores the physics of strong
coherent coupling between local systems (eg quantum dots\cite{Blick98}), quantum limited measurements (eg. using single
electronics\cite{Schoen97,Gurvitz97,Schoen98}), and proposals for quantum computation\cite{Kane98,Loss98}.

Our treatment will be based on a particularly simple system; a
single quantum dot coupled to two quantum channels, see figure \ref{fig1}, as this system is the electronic analogue of a
single Fabry-Perot cavity in quantum optics. The conductance properties of this system can easily be obtained by the method of
Buttiker\cite{Buttiker}. Recently a similar result was obtained using a Markov master equation description of
the quasibound state of the dot\cite{Sun99}. In this paper we give an equivalent description in terms of the quantum
stochastic calculus for Fermi fields. 

\section{Quantum stochastic calculus for Fermions}
The free Hamiltonian for a Fermion channel is 
\begin{equation}
H=\hbar\sum_k \omega_k a_k^\dagger a_k
\end{equation}
where $a_k$ is a Fermi annihilation operator satisfying the anti commutation relations
\begin{eqnarray}
a_k a_l+a_la_k & = & 0\\
a_k a_l^\dagger+a_l^\dagger a_k & = & \delta_{kl}
\end{eqnarray}
We now define the Fermi field operator
\begin{equation}
a(t)=\sum_ka_ke^{-i(\omega_k-\omega_0)t}
\end{equation}
which represents free field dynamics with respect to  a frame rotating at frequency $\omega_0$. This frequency will later be taken
to characterise the energy of the quasi bound state to which the Fermi field is coupled. The field operator
satisfies the continuum anti commutation relations
\begin{equation}
a(t)a^\dagger(t^\prime)+a^\dagger(t^\prime)a(t)=\delta(t-t^\prime)
\end{equation}
Our objective is to define a quantum stochastic process to accurately characterise the Fermi statistics of these fields. To
that end we define the integrated operators,
\begin{equation}
A(t)=\int_0^t dt^\prime a(t^\prime)
\end{equation}
and the corresponding Ito increment
\begin{equation}
dA(t)=A(t+dt)-A(t)
\end{equation}
We now need to specify the state of the free fields. We will take these to be thermal equilibrium states of a noninteracting
Fermi system at temperature $T$. It is then easy to show, under appropriate assumptions that
\begin{eqnarray}
\langle dA^2\rangle & = & \langle (dA^\dagger)^2\rangle  =  0\\
\langle dA^\dagger dA\rangle & = & f(\omega_0)dt\\
\langle dA dA^\dagger\rangle & = & (1-f(\omega_0))dt
\end{eqnarray}
where the equilibrium occupation probability $f(E)$ is evaluated at the reference energy $E_0=\hbar\omega_0$. Later this will
be the probability that a free field Fermi state, resonant with quasi bound state, is occupied. The important point to note
here is that these quantities, while quadratic in the field increments, are only first order in the time increment. This is a
quantum analogue of the classical Wiener stochastic process\cite{Gardiner91}. Stochastic integrals of averaged field operators
are found using a generalisation of the Ito calculus for classical stochastic processes\cite{Gardiner83}. In particular we
have,
\begin{eqnarray}
\int_0^{t_1} dt\int_0^{t_2} dt^\prime \langle dA^\dagger(t)dA(t^\prime)\rangle
X(t)Y(t^\prime) &= &\int_0^{min(t_1,t_2)}dtf(\omega_0)X(t)Y(t)\\
\int_0^{t_1} dt\int_0^{t_2} dt^\prime \langle dA(t)dA^\dagger(t^\prime)\rangle
X(t)Y(t^\prime) &= &\int_0^{min(t_1,t_2)}dt(1-f(\omega_0))X(t)Y(t)
\label{ito-rule}
\end{eqnarray}
where $X(t),Y(t)$ may be operator valued functions. We also note that in the Ito calculus the chain rule is modified
according to\cite{Barchielli86}
\begin{equation}
d(A(t)B(t))=dA(t)B(t)+A(t)dB(t)+dA(t)dB(t)
\end{equation}
The final term must be included to ensure a correct expansion to linear order in the time increment.

We need now to specify how the free field is coupled to a local electronic degree of freedom described by the Fermi
annihilation and creation operators $c,c^\dagger$. This degree of freedom could for example be a quasi bound state of a single
quantum dot or donor atom.  This interaction will be taken to be linear in the system operators. 
The usual tunnelling interaction is then specified by the Hamiltonian,
\begin{equation}
H_I=\sum_k g_k a^\dagger_k c+g^*_k a_k c^\dagger
\end{equation}
Such a coupling is the usual way to describe tunnel coupling between a reservoir and a localised degree of
freedom\cite{tunnel-coupl} We now anticipate the Markov approximation by assuming that around the reference energy, the
coupling constants, $g_k$ are very slowly varying functions of $k$ and replace them by a constant, $\sqrt{\gamma}$. For further
discussion see\cite{ColGar}. 
\begin{equation}
H_I=\sqrt{\gamma}(ca^\dagger(t)+c^\dagger a(t))
\end{equation} 

The time evolution operator over the time increment $dt$ is given by
\begin{equation}
U(dt)=\exp\{-i\sqrt{\gamma}(cdA^\dagger(t)+c^\dagger dA(t))\}
\end{equation}
This enables us to define the 'output' field stochastic process as 
\begin{equation}
dA_{out}(t)= U^\dagger(dt)dA(t)U(dt)
\end{equation}
This equation suggests that we regard $dA(t)$ as the 'input' field stochastic process $dA(t)=dA_{in}(t)$. It is then easy to
see that
\begin{equation}
dA_{out}(t)=dA_{in}(t)-i\sqrt{\gamma}(2f(\omega_0)-1)c(t)dt
\end{equation}
This expression may be written directly in terms of the reservoir field operators as
\begin{equation}
a_{out}(t)=a_{in}(t)-i\sqrt{\gamma}(2f(\omega_0)-1)c(t)
\end{equation}
This expression can be used to establish a connection between input and output fields once
the dynamics of the local system operators is given. In the frequency (energy) domain the resulting expression is equivalent
to the scattering matrix in the method of Buttiker. 

The dynamics of the local system is specified by a quantum stochastic differential equation, for example
\begin{eqnarray}
dc(t) & = & U^\dagger(dt)c(t)U(dt)-c(t)\\
& = & -\frac{\gamma}{2}c(t)+i\sqrt{\gamma}dH(t)
\end{eqnarray}
where the noise operators are 
\begin{equation}
dH(t)=dA(t)(c(t)c^\dagger(t)-c^\dagger(t) c(t))
\label{sde}
\end{equation}
It is at this point that we recognise the difference between Fermi and Bose quantum stochastic calculus. In the Bose case the
noise operator does not depend on system operators, and is in fact simply given by $dA(t)$, as the commutator in Eq \ref{sde}
is unity. In the Fermi case however {\em the noise  depends on the system operators}. This will ensure that the dynamics of the
local system  reflects Fermi statistics, in particular it will ensure that the stochastic differential equation for the
number operator
$n_c(t)=c^\dagger(t)c(t)$ takes the correct form. Despite the fact that the noise  operators depend on the system operators,
the average over quadratic combinations do not depend on the system. This is because the system operator appearing in
Eq(\ref{sde}) has only two eigenvlaues $\pm1$, which when squared is unity. Thus we find
\begin{equation}
dH^\dagger(t) dH(t)=dA^\dagger(t) dA(t)
\end{equation}

The quantum stochastic differential equation for the system number operator is given by 
\begin{eqnarray}
dn_c(t) & = & c^\dagger(t)dc(t)+dc^\dagger(t) c(t)+dc^\dagger(t)dc(t)\\
 & = & -\gamma (n_c(t)-f(\omega_0))dt+dN(t)
\label{num-qsde}
\end{eqnarray}
where the noise operator is 
\begin{equation}
dN(t)=i\sqrt{\gamma}(c(t)^\dagger dH(t)-c(t)dH^\dagger(t))
\end{equation}
Equation \ref{num-qsde} correctly reflects the Fermi statistics of the local system. If the Ito correction term,
$dc^\dagger(t)dc(t)$ had been neglected, this would not have occurred. When this term is evaluated, the dependence of the
noise operators on system variables  in Eq. \ref{sde} is crucial.   As a result the average occupation of the local system in
the steady state is given correctly by  
\begin{equation}
\langle n_c\rangle=f(\omega_0)
\end{equation}
as would be expected for a Fermi particle.  

We can also obtain the master equation for the local system by
\begin{equation}
d\rho(t)=\mbox{Trace}_{\cal R}\left [U(dt)W(t)U^\dagger(dt) -W(t)\right ]
\end{equation}
where $W(t)$ is the state of the total system (localised system and the external fields), and Trace refers to a trace over
external field variables. Using the noise moments we find,
\begin{eqnarray}
\frac{d\rho(t)}{dt}  & = & -i[H,\rho(t)]+\gamma f(\omega_0)(2c^\dagger\rho c-cc^\dagger\rho(t)-\rho(t)cc^\dagger)\\ \nonumber
& & \mbox{}+\gamma(1- f(\omega_0))(2c\rho c^\dagger-c^\dagger c\rho(t)-\rho(t)c^\dagger c)
\end{eqnarray}
It is an easy matter to verify that the mean occupation of the dot $\langle n_c(t) \rangle$ obeys the same equation that 
results from taking moments of both sides of Eq. \ref{num-qsde}. The master equation represents the Schr\"{o}dinger picture
dynamics, while the quantum stochastic differential represents the Heisenberg picture.

\section{Quantum stochastic dynamics of a single quantum dot}
The system we will discuss in this paper is  a standard mesoscopic configuration\cite{datta} in which ohmic contacts couple to
propagating channels to either side of a quantum dot (see figure \ref{fig1}). We suppose that there  is a single quasi bound
state between two tunnel barriers. Spin will be ignored. It can easily be included as another  state in the dot. We will also
ignore Coulomb blockade, which for a single bound state simply leads to a shift in the energy of the state. We  also take
the so called 'zero-temperature' limit, and assume that the bound state energy is below the effective Fermi energy in the
source (L, in figure \ref{fig1}) and above the effective Fermi energy in the drain (R in figure
\ref{fig1}).  The ohmic contact at the left of the dot is assumed to be a perfect emitter while the ohmic contact at the right
of the dot is a perfect absorber. This ensures that both  reservoirs will remain close to thermal equilibrium at all times,
provided they are connected to an external EMF. These assumptions will be important when we consider the measured quantities in
this system. 

We need now to specify how the free fields in the left and right channels are coupled to a local electronic degree of freedom
described by the Fermi annihilation and creation operators $c,c^\dagger$. This degree of freedom could for example be a quasi
bound state of a single quantum dot or donor atom.  This interaction will be taken to be linear in the system operators. 
The usual tunnelling interaction, anticipating the Markov approximation as in the previous section, is then specified by the
Hamiltonian,
 \begin{equation}
H_I=\sqrt{\gamma_L}(ca^\dagger_L(t)+c^\dagger a_L(t))+\sqrt{\gamma_R}(ca^\dagger_R(t)+c^\dagger a_R(t))
\end{equation}
where $\gamma_L,\gamma_R$ refers to the tunnelling rate across the left and right barrier respectively, while $a_L(t),a_R(t)$
are the Fermi fields in the left channel and right channel respectively. The corresponding  unitary
evolution operator for a time increment $dt$ is 
\begin{equation}
U(dt)=\exp\left (-i\sqrt{\gamma_L}(cdA_L^\dagger+c^\dagger dA_L)-i\sqrt{\gamma_R}(cdA_R^\dagger+c^\dagger dA_R)\right )
\end{equation}
where $dA_L(t),dA_R(t)$ are the quantum stochastic processes in the left and right channels respectively. The
input/output relations are then found to be
\begin{eqnarray}
dA_{L,out}(t) & = & dA_{L,in}(t)-i\sqrt{\gamma_L}c(t)dt\\
dA_{R,out}(t) & = & dA_{R,in}(t)+i\sqrt{\gamma_R}c(t)dt
\end{eqnarray}
with the following averages for the noise 
\begin{eqnarray}
\langle dA_L(t)^\dagger dA_L(t)\rangle & = & dt\\
\langle dA_R(t) dA_R^\dagger(t)\rangle & = & dt
\end{eqnarray}
All other averages are zero. Alternatively we may write the input-output relations as
\begin{eqnarray}
 a_{L,out}(t) & = & da_{L,in}(t)-i\sqrt{\gamma_L}c(t)\\
a_{R,out}(t) & = & a_{R,in}(t)+i\sqrt{\gamma_R}c(t)
\label{input-output}
\end{eqnarray}

The quantum stochastic differential equation for the destruction operator in the dot is,
\begin{equation}
dc(t)=-\frac{(\gamma_L+\gamma_R)}{2}c(t)dt+i\sqrt{\gamma_L}dH_L(t)+i\sqrt{\gamma_R}dH_R(t)
\end{equation}
The quantum stochastic differential equation for the number operator on the dot is
\begin{equation}
dn_c(t)=\gamma_L(1-n_c(t))dt-\gamma_Rn_c(t)dt+dN_L(t)+dN_R(t)\ .
\label{dot-qsde}
\end{equation}
If we take moments of both sides of Eq.\ref{dot-qsde} we find,
\begin{equation}
\frac{d\bar{n}}{dt} = \gamma_L(1-\bar{n})-\gamma_R\bar{n}
\end{equation}
The first term corresponds to injection from the source onto the dot. This term is zero if the dot is already occupied
and $\bar{n}(t)=1$. The second term correspond to emission from the dot through the right barrier into the drain. The steady
state occupation number on the dot is 
\begin{equation}
\bar{n}_\infty=\frac{\gamma_L}{\gamma_L+\gamma_R}
\end{equation}
The master equation for the dot is found to be
\begin{eqnarray}
\frac{d\rho}{dt}={\cal L}\rho & = & \frac{\gamma_L}{2}\left (2c^\dagger \rho c-cc^\dagger \rho-\rho cc^\dagger\right )\\
\nonumber
		          & &\mbox{}+\frac{\gamma_R}{2}\left ( 2c\rho c^\dagger -c^\dagger c \rho-\rho c^\dagger c\right )
\label{master_eq}
\end{eqnarray}
This equation was previously derived by more direct methods in reference \cite{Sun99}.

\section{What is measured ?} 

It is at this point we need to make contact with measurable quantities. In the case of electron transport, the measurable
quantities reduce to current $I(t)$ and voltage $V(t)$. The measurement results are a time series of currents and voltages which
exhibit both systematic and stochastic components. Thus $I(t)$ and voltage $V(t)$ are classical conditional stochastic
processes, conditioned by the underlying quantum dynamics of the quasi bound state on the dot. The reservoirs in the ohmic
contacts play a key role in defining the measured quantities and ensuring that they are ultimately classical stochastic
processes. Transport through the dot results in charge fluctuations in either the left or the right channels. These
fluctuations decay extremely rapidly, ensuring that the channels remain in thermal equilibrium with the respective ohmic
contacts. For this to be possible charge must be able to flow into and out of the channels from an external circuit. We
assume that a constant potential difference is maintained between the two reservoirs either side of the dot. While the entire
system is clearly not in thermal equilibrium, we assume that the left and right channels are themselves close to thermal
equilibrium and can each be specified by a separate chemical potential
$\mu_L$ and $\mu_R$, and these are held constant by a external voltage source, $V$. 

If a single electron tunnels out of the dot into the right channel between time $t$ and $t+dt$, its energy is momentarily
above the Fermi energy. This electron scatters very strongly from the electrons in that channel and propagates into the right
ohmic contact where it is perfectly absorbed. The net effect is a small current pulse,
$dI_L(t)$ in the external circuit. This is completely analogous to perfect photodetection: a photon emitted from a cavity will
be detected with certainty by a detector which is a perfect absorber.  Likewise when an electron in the right channel tunnels
onto the dot, there is a rapid relaxation of this unfilled state back to thermal equilibrium as an electron is emitted from the
right ohmic contact into the depleted channel. This again results in a current pulse in the  circuit connected to the ohmic
contacts. The energy gained  when one electron is emitted from the left reservoir is, by definition, the chemical potential of
that reservoir,
$\mu_L$ while the energy lost when one electron is absorbed into the right reservoir is $\mu_R$. The net energy transferred
between reservoirs is $\mu_L-\mu_R$. This energy is supplied by the external EMF, $V$  and thus $\mu_L-\mu_R=eV$. It should not
be supposed that the electron injected from the left contact and emitted into the right contact have the same energy as the
energy of an electron on the dot. In fact any  electron energy at all will suffice to restore thermal equilibrium in the left
and right leads. If an electron is emitted into the left channel between times $t$ and $t+dt$, the (unnormalised) sate of that
channel is $\rho_{L,O}(t+dt)=a_{L,O}^\dagger\rho(t)a_{L,O}dt$. The probability of this event occurring is simply the
normalisation constant and is $p_e(t)=\mbox{tr}(a_{L,O}^\dagger\rho(t)a_{L,O}) dt=\langle
a_{L,O}(t)a_{L,O}^\dagger(t)\rangle$. That is to say the probability of emission of electrons into the left channel is
determined by the {\em anti normally ordered} number flux operator in the left channel. A similar argument indicates that the
probability of absorption of an electron in the right ohmic contact is given by the mean of the {\em normally ordered} number
flux operator in the right most channel,
$p_a(t)=\langle a_{R,O}^\dagger(t)a_{R,O}(t)\rangle$ . 
This is precisely analogous to perfect photodetection from an optical source\cite{WallsMilb}. Both the emission into the left
channel and absorption from the right channel are conditional point processes, conditioned on the quantum state of the quasi
bound state on the dot. 

On average of course the same
current flows in both reservoirs, however as the current is stochastic it is made up of contributions from pulses in each
lead, which do not necessarily occur at the same time. Indeed they  are necessarily separated in time by a degree
depending on the life time of the quasi bound state in the dot.  In mesoscopic devices however current in measured locally in
each lead, thus we can consider either the current in the left lead, $I_L(t)$ or the current in the right lead $I_R(t)$ and
correlations between them.   The
current that flows in the right lead  is simply given by the probability per unit time that an electron in that channel
is absorbed by the perfect absorber that is the right Ohmic contact. Thus
\begin{equation}
E(I_R(t))=e\langle a_{R,O}^\dagger(t)a_{R,O}(t)\rangle
\end{equation}
The average current that flows in the left lead is given by the average probability per unit time that an electron is emitted
by the perfect emitter that is the left Ohmic contact.
\begin{equation}
E(I_L(t))=e\langle a_{L,O}(t)a_{L,O}^\dagger(t)\rangle 
\end{equation}
Note that the average on the left hand side is an average of a classical stochastic process, while the average on the right is
of a quantum stochastic process. 
We may now substitute the relationship between the output fields, the input fields and the operator for the local state,
Eqs(37,38 ). The average currents are then found to be
\begin{eqnarray} 
E(I_L(t)) & = & e\gamma_L(1-\langle n_c(t)\rangle)\\
E(I_R(t)) & = & e\gamma_R\langle n_c(t)\rangle
\end{eqnarray}
 In the stationary state both currents are equal and given by
\begin{equation}
I_{L,\infty}=I_{R,\infty}=\frac{e\gamma_L\gamma_R}{\gamma}
\end{equation}
where $\gamma=\gamma_L+\gamma_R$. 

The stationary two time correlation matrix is  given by
\begin{equation}
G_{\alpha,\beta}(\tau)=E(I_\alpha(t+\tau),I_\beta(t))_{t\rightarrow\infty}
\end{equation}
and $E(X,Y)=E(XY)-E(X)E(Y)$. 
The quantity $E(I_\alpha(t+\tau)I_\beta(t)$ is determined by the appropriately ordered two time correlation function for the
quantum fields in the channels. As both emission and absorption are point processes we find,
\begin{eqnarray}
 E(I_L(t+\tau)I_L(t) & = & e^2\langle a_{L,O}(t)a_{L,O}^\dagger(t)\rangle\delta(\tau)+e^2\langle
a_{L,O}(t)a_{L,O}(t+\tau)a_{L,O}^\dagger(t+\tau)a_{L,O}^\dagger(t)\rangle_{\tau > 0}\\
E(I_R(t+\tau)I_R(t)) & = & e^2\langle a_{R,O}(t)^\dagger a_{R,O}(t)\rangle\delta(\tau)+e^2\langle
a^\dagger_{R,O}(t)a^\dagger_{R,O}(t+\tau)a_{R,O}(t+\tau)a_{R,O}(t)\rangle_{\tau >0}\\
E(I_R(t+\tau)I_L(t)) & = & \langle a^\dagger_{R,O}(t+\tau)a_R(t+\tau) a_{L,O}(t)a_{L,O}^\dagger(t)\rangle\\
E(I_L(t+\tau)I_R(t)) & = & \langle a_{L,O}(t+\tau)a^\dagger_{L,O}(t+\tau)a_{R,O}^\dagger(t)a_{R,O}(t)\rangle
\end{eqnarray}

Using equations 37 and 38, the steady state $(t\rightarrow\infty$) field correlation functions  may be expressed solely in
terms of the correlation functions for the quasibound state as
\begin{eqnarray}
\langle a_{L,O}(t)a_{L,O}(t+\tau)a_{L,O}^\dagger(t+\tau)a_{L,O}^\dagger(t)\rangle_{\tau > 0} & = & \gamma_L^2 \mbox{Tr} \left
(cc^\dagger e^{{\cal L}\tau}c^\dagger\rho_\infty c\right )\\
\langle a^\dagger_{R,O}(t)a^\dagger_{R,O}(t+\tau)a_{R,O}(t+\tau)a_{R,O}(t)\rangle_{\tau >0} & = & \gamma_R^2\mbox{Tr}\left (
c^\dagger ce^{{\cal L}t\tau}c\rho_{\infty}c^\dagger\right )\\
\langle a^\dagger_{R,O}(t+\tau)a_R(t+\tau) a_{L,O}(t)a_{L,O}^\dagger(t)\rangle & = & \gamma_R\gamma_L\mbox{Tr}
\left(c^\dagger c e^{{\cal L}\tau}c^\dagger \rho_\infty c\right )  \\ 
\langle a_{L,O}(t+\tau)a^\dagger_{L,O}(t+\tau)a_{R,O}^\dagger(t)a_{R,O}(t)\rangle & = & \gamma_L\gamma_R\mbox{Tr}\left
(cc^\dagger e^{{\cal L}\tau} c\rho_\infty c^\dagger\right )
\end{eqnarray}
where $\rho_\infty$ is the steady state solution to the master equation, Eq \ref{master_eq} for the quasi bound state on the
dot. 

Upon solving the master equation we may evaluate each correlation function to give;
\begin{eqnarray}
 E(I_L(t+\tau),I_L(t)  & = & E(I_R(t+\tau),I_R(t)
=e^2\frac{\gamma_L\gamma_R}{\gamma}\delta(\tau)-e^2\frac{\gamma_R^2\gamma_L^2}{\gamma^2}e^{-\gamma \tau}\\ E(I_R(t+\tau),I_L(t)
& = & e^2\frac{\gamma_L\gamma_R^3}{\gamma^2}e^{-\gamma \tau}\\ E(I_L(t+\tau),I_R(t) & = &
e^2\frac{\gamma_R\gamma_L^3}{\gamma^2}e^{-\gamma \tau}
\end{eqnarray}
where $\gamma=\gamma_L+\gamma_R$. 
The power spectrum of the noise in either the left or the right lead is then given by the Fourier transform of the current
correlation function in either lead
\begin{equation}
S_{R}(\omega)=S_L(\omega)= ei_\infty\left (1-\frac{2\gamma_R\gamma_L}{\gamma^2+\omega^2}\right )
\end{equation}
Note that at zero frequency and symmetric rates ($\gamma_R=\gamma_L$) the current noise is suppressed by a factor of 0.5 over
shot noise.  This is the same as that obtained by Buttiker\cite{Buttiker92}, and is equivalent to that obtained in reference
\cite{Sun99}. Note however that in that paper it was  assumed, following earlier work\cite{Chen91}, that the measured current
is a superposition of the two Poisson processes of emission and absorption. This however is not the case for mesoscopic
measurements. As has been stressed by Buttiker, the current is measured in one lead at a time and thus the correct expression
for the current noise is that given above. 

\section{discussion and conclusion}
In this paper we have shown an equivalence between the approach of Buttiker and the approach of quantum stochastic calculus to
the current through mesoscopic systems. To illustrate the equivalence we have discussed the current fluctuations in two
terminal mesoscopic circuit with two tunnel barriers containing a single quasi bound state on the well. The method enables us
to focus on either the incoming and outgoing Fermi fields in the leads, or on the irreversible dynamics of the well state
itself.  We have of course made the Markov approximation in order to obtain the quantum Langevin equations for the system. The
Markov assumption is equivalent to the assumption of the Breit-Wigner (Lorentzian) assumption for the transmission coefficients
though a double barrier structure, and is discussed in some detail in reference \cite{Sun99}. 

We believe there are two advantages in our approach. Firstly it is useful to be able to refer to the quantum irreversible
dynamics of the quasi bound states on local systems defined by barriers, as well as the input and output Fermi fields. This is
particularly important in coherently coupled quasibound states. This is essential for  recent condensed matter schemes for quantum
computation, where the focus is not so much on the properties of the external, classical currents but rather on the dynamics of
the local systems themselves. Secondly, our method parallels a similar approach to strongly coupled field modes in quantum
optics, thus suggesting useful directions for future work. We mention one such direction. In quantum optics the method of
quantum trajectories\cite{Knight98} enables a description to be given of quantum limited measurement\cite{Carmichael-book},
quantum control (feedback and feed forward)\cite{Wiseman93}, and cascaded local systems (irreversible but directional
coupling)\cite{Carmichael-cascade}. As mesoscopic technology advances these topics will become increasingly important. The
presentation in this paper shows how the quantum stochastic methods of quantum optics may be taken over to Fermi systems.

\acknowledgements
This work was supported by the SP group, Cavendish Laboratory, Cambridge University. I would like to thank Crispen Barnes,
Chris Ford and Matthew Collett for useful discussions.

\begin{figure}
\caption{Schematic representation of tunneling through a single quantum dot.}
\label{fig1}
\end{figure}

\end{document}